\documentclass{article}
\usepackage{appendix}
\PassOptionsToPackage{numbers,sort&compress}{natbib}

\usepackage[preprint]{neurips_2024} 
\usepackage{authblk} 
\usepackage{amsmath}
\usepackage{units}
\usepackage{xspace}

\newcommand{\github}[1]{\href{https://github.com/#1}{\adjustbox{valign=c}{\includegraphics[width=8pt]{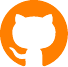}}}}

\usepackage[utf8]{inputenc} 
\usepackage[T1]{fontenc}    
\usepackage{url}            
\usepackage{booktabs}       
\usepackage{amsfonts}       
\usepackage{nicefrac}       
\usepackage{microtype}      
\usepackage{xcolor}         
\usepackage{ulem}
\usepackage{adjustbox}      
\usepackage{subcaption}

\usepackage{hyperref}   
\hypersetup{
  colorlinks   = true, 
  urlcolor     = blue, 
  linkcolor    = blue, 
  citecolor    = blue 
}

\usepackage[disable]{todonotes}  

\title{FAIR Universe\\Weak Lensing ML Uncertainty Challenge:\\
Handling Uncertainties and Distribution Shifts for Precision Cosmology}

\author[1\textbf{*}]{Biwei Dai}
\author[2\textbf{*}]{Po-Wen Chang}
\author[2]{Wahid Bhimji}
\author[2]{Paolo Calafiura}
\author[3,4]{Ragansu Chakkappai}
\author[5]{Yuan-Tang Chou}
\author[2]{Sascha Diefenbacher}
\author[6,2]{Jordan Dudley}
\author[7,2] {Ibrahim Elsharkawy}
\author[2]{Steven Farrell}
\author[4]{Isabelle Guyon}
\author[2]{Chris Harris}
\author[8,5,2]{Elham E Khoda}
\author[9,10]{Benjamin Nachman}
\author[3,4]{David Rousseau}
\author[6,2]{Uroš Seljak}
\author[4]{Ihsan Ullah}
\author[5]{Yulei Zhang}

\affil[1]{Institute for Advanced Study}
\affil[2]{Lawrence Berkeley National Laboratory}
\affil[3]{Universit\'e Paris-Saclay, CNRS/IN2P3, IJCLab}
\affil[4]{ChaLearn}
\affil[5]{University of Washington}
\affil[6]{University of California, Berkeley}
\affil[7]{University of Toronto}
\affil[8]{University of California, San Diego}
\affil[9]{Stanford University}
\affil[10]{SLAC National Accelerator Laboratory}
\affil[*]{\textbf{Equal contribution} (with remaining names in alphabetical order)}

\date{}

\begin{document}

\maketitle

\begin{center}
    \texttt{\textcolor{blue}{fair-universe@lbl.gov}} \\
    \url{https://fair-universe.lbl.gov} \\~\\
    \today
\end{center}

\begin{abstract} 
Weak gravitational lensing, the correlated distortion of background galaxy shapes by foreground structures, is a powerful probe of the matter distribution in our universe and allows accurate constraints on the cosmological model. 
In recent years, high-order statistics and machine learning (ML) techniques have been applied to weak lensing data to extract the nonlinear information beyond traditional two-point analysis. 
However, these methods typically rely on cosmological simulations, which poses several challenges: simulations are computationally expensive, limiting most realistic setups to a low training data regime; inaccurate modeling of systematics in the simulations create distribution shifts that can bias cosmological parameter constraints; and varying simulation setups across studies make method comparison difficult.
To address these difficulties, we present the first weak lensing benchmark dataset with several realistic systematics and launch the FAIR Universe Weak Lensing Machine Learning Uncertainty Challenge. The challenge focuses on measuring the fundamental properties of the universe from weak lensing data with limited training set and potential distribution shifts, while providing a standardized benchmark for rigorous comparison across methods. Organized in two phases, the challenge will bring together the physics and ML communities to advance the methodologies for handling systematic uncertainties, data efficiency, and distribution shifts in weak lensing analysis with ML, ultimately facilitating the deployment of ML approaches into upcoming weak lensing survey analysis.
\end{abstract}

\clearpage
\tableofcontents

\clearpage

\section{Introduction}

\subsection{Background and impact}
\label{sec:background} 

Modern science increasingly relies on large and complex datasets measured by advanced instruments, a trend recently accelerated by machine learning (ML) and artificial intelligence (AI). To make robust discovery with high-dimensional data, we need to develop efficient methods that can accurately quantify uncertainties (particularly systematic or epistemic uncertainties) and address the potential bias from our inaccurate modeling of the physical systems (also known as distribution shifts). Cosmology, specifically the analysis of the large-scale structure of the universe, offers a compelling case study of these challenges in the pursuit of fundamental physical insights.


The large-scale structure of the universe, i.e., the cosmic web of galaxies, galaxy clusters, and dark matter spanning hundreds of millions of light-years, encodes essential information about the composition, evolution, and fundamental laws governing the cosmos. However, the majority of matter in the universe is dark matter, which does not interact with light and can only be observed indirectly through its gravitational effects. According to Einstein’s theory of general relativity, the gravitational field of this large-scale structure bends the path of light traveling through the universe. Weak gravitational lensing refers to the subtle, coherent distortions in the observed shapes of distant galaxies caused by the deflection of light as it traverses the inhomogeneous matter distribution of the universe. By statistically analyzing these distortions across large regions of the sky, weak lensing provides a powerful probe of the matter distribution and the underlying cosmological model~\cite{hoekstra2008a, kilbinger2015a}.

Traditional analysis based on two-point correlation functions can only capture a limited amount of information from the weak lensing data (which can be postprocessed to 2D fields similar to images). To fully exploit the non-Gaussian features present in the cosmic web, higher-order statistics and modern ML methods have become increasingly important \cite{lanusse2023dawes}. These approaches, including deep learning and simulation-based inference \cite{cranmer2020frontier}, have been shown to extract significantly more information in weak lensing maps than traditional techniques \cite{fluri2018cosmological,Gupta2018a,dai2022translation,lu2023cosmological,dai2024multiscale,sharma2024comparative,cheng2025cosmological,jeffrey2025dark,von2025kids,zeghal2025simulation}. However, different analyses assume different dataset setups and lead to different results, making it hard to directly compare with existing approaches. Furthermore, most (if not all) of these methods rely heavily on simulations that may not accurately represent real data due to modeling approximations and missing systematics \cite{villaescusa2021multifield}. 

Motivated by the need to quantify and compare the information content that different analysis methods can extract from weak lensing maps, while also evaluating their robustness to simulation inaccuracies and observational systematics, we launch the Weak Lensing ML Uncertainty Challenge. 
This data challenge is designed to benchmark a broad class of approaches, including higher-order statistics, machine learning algorithms, and simulation-based inference techniques, under realistic conditions. Participants will analyze a suite of carefully designed mock weak lensing maps with known cosmological parameters, constructed to include variations in simulation fidelity and observational systematics.

The challenge is organized into two phases of machine learning competitions, with particular emphasis on handling biased (i.e., systematic-contaminated) test data. Participants are required not only to provide point estimates of $(\Omega_m, S_8)$, but also to quantify the associated uncertainties. In addition, to probe robustness against simulation-model mismatch, some test datasets are generated from alternative physical models. Participants must therefore perform \textit{out-of-distribution} (OoD) detection to identify data that are inconsistent with the training simulations. By comparing the performance and robustness of different methods in a controlled setting, the challenge aims to systematically assess their ability to extract cosmological information while quantifying their sensitivity to modeling assumptions and systematics.

To join the challenge, participants have to sign up for either the \href{https://www.codabench.org/competitions/8934/}{Phase-1 competition} (ended in November 2025\footnote{The final results and the top solutions developed by our participants are available on \href{https://fair-universe.lbl.gov/WeakLensing-Uncertainty-Challenge.html}{our website} and our \href{https://fair-universe.lbl.gov/WeakLensing-Uncertainty-Challenge-Workshop.html}{NeurIPS 2025 workshop agenda}.}) and the \href{https://www.codabench.org/competitions/10902/}{Phase-2 competition} (open for submission until October 2026) websites on Codabench~\cite{XU2022100543}, an open-source platform for data science benchmarks. For more information, please check out our \href{https://github.com/FAIR-Universe/Cosmology_Challenge}{GitHub repository}~\github{FAIR-Universe/Cosmology_Challenge}. 

The outcomes of this challenge are expected to guide the development of next-generation weak lensing analysis pipelines, foster cross-disciplinary collaboration between the astrophysics and machine learning communities, and ultimately improve the reliability of cosmological inference from current and upcoming surveys such as \href{https://sci.esa.int/web/Euclid}{Euclid Space Telescope},  
the \href{https://lsst.org}{Vera Rubin Observatory}, or  \href{https://roman.gsfc.nasa.gov/}{Nancy Grace Roman Space Telescope}. By explicitly addressing simulation-model mismatch and the need to quantify systematic uncertainties, this challenge emphasizes scientific robustness and interpretability, aligning with the growing emphasis on trustworthy ML in scientific domains.

\subsection{Novelty}
\label{sec:novelty}
To our knowledge, this is the first ML data challenge on realistic weak gravitational lensing datasets that explicitly addresses systematic uncertainties and distribution shifts. The ML competition is built on our experience running several competitions in particle physics and beyond. These include the FAIR Universe HiggsML uncertainties challenge \cite{Benato:2024lnj} (NeurIPS 2024 competition), the TrackML Challenge \cite{TrackMLAccuracy2019} (NeurIPS 2018 competition), the LHC Olympics \cite{Kasieczka:2021xcg}, AutoML/AutoDL (NeurIPS 2019 competition) \cite{liu2021tpami}, and other competitions \cite{elbaz2021pmlr,carriónojeda2022neurips22, guyon2012analysis, crowd-bias-challenge, ccaiunict-2023}, but none of them tackled the realistic tasks using the high-dimensional cosmological dataset.

\section{Data}

\begin{table}[tbp]
\centering
\begin{tabular}{c|ccccc}
\hline
redshift & box size & $N_{\mathrm{particle}}$ & force resolution & $N_{\mathrm{step}}$ & simulation code\\
\hline
$0<z<0.45$ & $320\ h^{-1}\mathrm{Mpc}$ & $960^3$ & $0.03\ h^{-1}\mathrm{Mpc}$ & adaptive & MP-Gadget \cite{yu_feng_2018_1451799} \\
$0.45<z<1.05$ & $704\ h^{-1}\mathrm{Mpc}$ & $2816^3$ & $0.125\ h^{-1}\mathrm{Mpc}$ & $60$ & FastPM \cite{Feng:2016yqz} \\
$1.05<z<2.72$ & $1536\ h^{-1}\mathrm{Mpc}$ & $1536^3$ & $0.5\ h^{-1}\mathrm{Mpc}$ & $15$ & FastPM \cite{Feng:2016yqz} \\
\hline
\end{tabular}
\caption{N-body simulation parameters for different redshift ranges.\label{tab:nbody}}
\end{table}

\begin{figure}[t]
    \centering
    \includegraphics[width=0.9\linewidth]{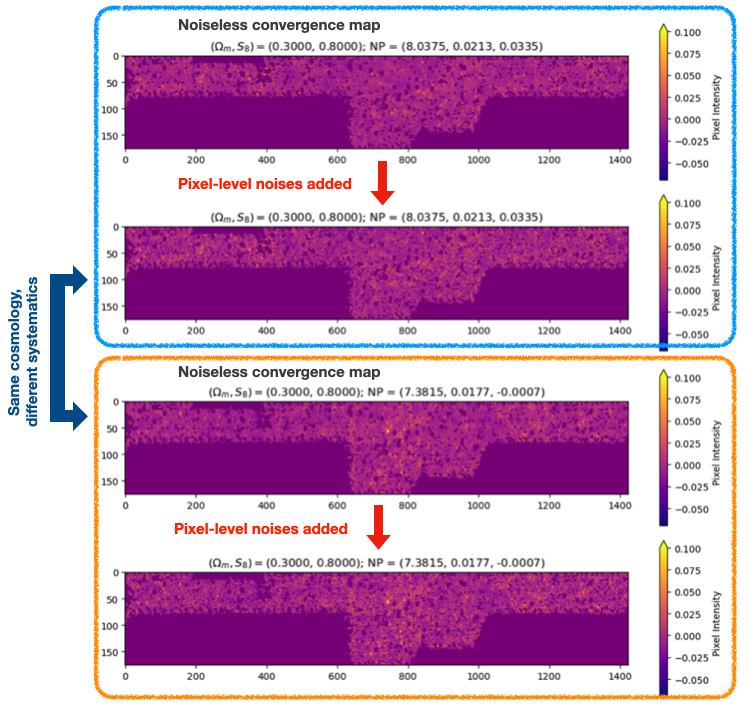}
    \caption{Example noiseless and noisy weak lensing convergence maps. The top panel and bottom panel show two different maps generated with different parameters of systematic effects and random seed, assuming the same cosmological parameters.}
    \label{fig:data_shift}
    \vspace{-0.2cm}
\end{figure}

We use a simulated weak gravitational lensing dataset for this competition to produce data representative of weak lensing measurement from the Hyper Suprime-Cam (HSC) survey \cite{Li:2021mvq}. The dataset is generated by running high-resolution cosmological N-body simulations \cite{yu_feng_2018_1451799,Feng:2016yqz} and a ray-tracing algorithm, assuming the same statistical properties (e.g., redshift distribution and survey mask) as the HSC Three-Year shape catalogue \cite{Li:2021mvq,rau2023weak}. Specifically, for each cosmological model, we run three N-body simulations with different box sizes and resolutions to cover the full redshift range and to meet the different volume and resolution requirements at different redshifts. The details of the simulation settings are summarized in Table \ref{tab:nbody}. $43$ snapshots (with a separation of $100 h^{-1}\mathrm{Mpc}$ between neighboring snapshots) are saved from these three simulation boxes for ray-tracing, and these snapshots are processed with the volume remapping technique \cite{carlson2010embedding} to cover the HSC Y3 footprint. Then we run multi-lens-plane ray-tracing algorithm \cite{jain2000ray, hilbert2009ray, Petri2016mocking} to simulate the weak lensing signals, under the flat-sky approximation. In data generation, we randomly translate, rotate, and flip the lens plane at each redshift to create pseudo-independent weak lensing maps \cite{petri2016sample}.

The original HSC Y3 galaxy catalog consists of 6 subfields on different parts of the sky, and are also divided into 4 redshift bins to enable tomographic analysis. For this data challenge, we only use the second redshift bin of the WIDE12H subfield to reduce the data size. The convergence maps (2D projections of the matter overdensity weighted by lensing geometry) with 2 $\mathrm{arcmin}$ resolution are estimated from these simulated galaxy catalogs as the dataset. Example convergence maps are shown in Figure \ref{fig:data_shift}. 

The weak lensing convergence maps are generated with $101$ different spatially-flat $\Lambda$CDM cosmological models. Each cosmological model differs in cosmological parameters $\Omega_m$, the fraction of the total matter density of the Universe, and $\sigma_8$, the amplitude of matter fluctuations on $8 \mathrm{Mpc}/h$ scales in the Universe today, while the other cosmological parameters are fixed at $\Omega_b=0.046$, $h=0.7$ and $n_s=0.97$. The two cosmological parameters $\Omega_m$ and $S_8=\sigma_8(\frac{\Omega_m}{0.3})^{0.5}$, shown in Figure \ref{fig:cosmology}, are sampled non-uniformly with density increases towards $\Omega_m = 0.3$ and $S_8 = 0.8$, to reduce the model interpolation error around fiducial cosmology. The parameter range is chosen to cover the posterior distributions from recent surveys including HSC Y3 \cite{li2023hyper,dalal2023hyper}, DES Y3 \cite{abbott2022dark}, and KiDS-1000 \cite{asgari2021kids}. These two parameters serve as the label of each data. 

In addition to the cosmological signal, we also model two of the most important systematic effects (distortions to the data), i.e., baryonic effect and photometric redshift uncertainty. The baryonic effect is modeled by applying a transfer function $T(k,z)$ to the simulated density field $\delta(k,z)$  to suppress the small-scale modes and mimic the AGN feedback \cite{sharma2025field}
\begin{equation}
    \delta(k,z) \rightarrow \delta(k,z)T(k,z),
\end{equation}
where $\delta(k)$ is the Fourier transform of the matter overdensity field $\delta(x) = \frac{\rho(x)}{\Bar{\rho}(x)}-1$, and $\rho(x)$ is the density map simulated from N-body simulations. The transfer function $T(k)$ is modeled using the halo model in HMcode \cite{mead2021hmcode}, with free parameters $T_{\mathrm{AGN}}$ and $f_0$ to parametrize the AGN feedback and star formation, respectively. These baryon parameters are sampled uniformaly in the prior range $T_{\mathrm{AGN}} \in [7.2,8.5]$, $f_0 \in [0,0.0265]$.
The photometric redshift uncertainty is modeled by shifting the source galaxy redshift distribution
\begin{equation}
    p(z) \rightarrow p(z+\Delta z),
\end{equation}
where $p(z)$ is the galaxy redshift distribution of the second redshift bin of HSC Y3, and the shift parameter $\Delta z$ is sampled from a Gaussian prior with mean $0$ and standard deviation $\sigma_{\Delta z}=0.022$ \cite{rau2023weak}.
These systematics are introduced in the data generation process, which we fully sampled in the training set so that the participants can marginalize over them. 
The free parameters in these systematic models $T_{\mathrm{AGN}}$, $f_0$, $\Delta z$ are nuisance parameters and need to be marginalized during inference. The measurement noise from intrinsic galaxy shapes is modeled as Gaussian noise with mean $0$ and standard deviation 
\begin{equation}
\sigma = \frac{\sigma_\epsilon}{\sqrt{2 n_{\rm \text{g}} A_{\rm \text{pixel}}}},
\end{equation}
where $\sigma_\epsilon \sim 0.4$ is the mean intrinsic ellipticity of galaxies, $A_{\rm \text{pixel}} = 4 ~\mathrm{arcmin}^2$ is the pixel area, and we set $n_g=30 ~\mathrm{arcmin}^{-2}$ to better probe the non-Gaussian information at small scales. Note that this noise level is lower than the current stage III weak lensing surveys and matches the upcoming cosmological surveys like Rubin Observatory LSST and Euclid. The shape noise can be added as a post-processing step during the training.

In total, we generate $256$ noiseless weak lensing maps with different realizations and different systematic parameters for each cosmological model. Each map has dimension $1424\times176$, so the total dimension of the training map is $101\times256\times1424\times176$. Each map corresponds to a unique 5D label $(\Omega_m, S_8, T_{\mathrm{AGN}},f_0,\Delta z)$, where the first two parameters are of interest, and the last three are nuisance parameters. 

Our simulated datasets incorporate major known systematics and are constructed to be as realistic as possible. As a result, we anticipate that models developed through this challenge will be applicable to real observational data \cite{Li:2021mvq}, enabling more robust and precise cosmological measurements. In particular, such measurements may shed light on the so-called $S_8$ tension \cite{amon2022non,de2025cosmic} — the current discrepancy between measurements of $S_8$ from early-universe observations and those from late-time large-scale structure, potentially revealing new physics beyond the standard cosmological model. Moreover, the insights gained from this challenge will be instrumental in informing the development of next-generation weak lensing analysis pipelines for upcoming surveys such as the \href{https://sci.esa.int/web/Euclid}{Euclid Space Telescope}, the \href{https://lsst.org}{Vera Rubin Observatory}, and  \href{https://roman.gsfc.nasa.gov/}{Nancy Grace Roman Space Telescope}.

\section{Tasks}

\begin{figure}[t]
    \centering
    \includegraphics[width=0.5\linewidth]{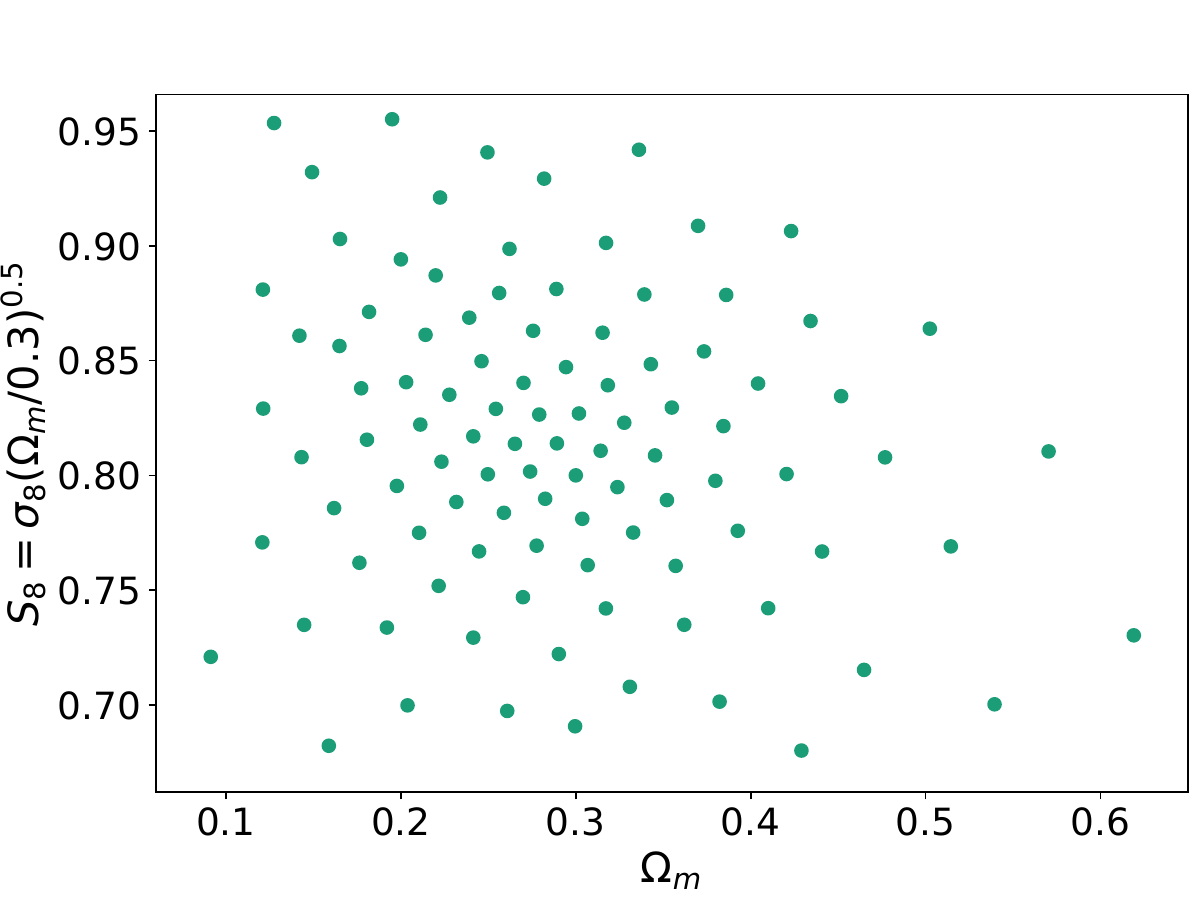}
    \caption{The $101$ cosmological parameters ($\Omega_m, S_8$) in our suite of weak lensing simulations.}
    \label{fig:cosmology}
    \vspace{-0.2cm}
\end{figure}

We design a two-phase competition to address both accurate cosmological parameter inference and robustness to distributional shifts in weak lensing analyses. The provided training dataset consists of $101 \times 256$ samples, where $101$ corresponds to distinct cosmological models spanning the $(\Omega_m, S_8)$ parameter space in Figure~\ref{fig:cosmology}, and each cosmological simulation is shifted by $256$ realizations of physical (nuisance) parameters. 

\subsection{Phase 1: Cosmological Parameter Inference}
In the first phase, participants are tasked with developing models to infer the cosmological parameters $(\hat{\Omega}_m, \hat{S}_8)$ and their associated one-standard-deviation uncertainties $(\hat{\sigma}_{\Omega_m}, \hat{\sigma}_{S_8})$ from simulated weak lensing convergence maps $\boldsymbol{x}$. These quantities may be obtained using, for example, posterior sampling methods such as Markov chain Monte Carlo (MCMC), or through maximum likelihood estimation.

The test data of this phase are drawn from the same distribution as the training data (i.e., \textit{in-distribution}, InD). 

\subsection{Phase 2: Out-of-Distribution Detection}
Due to limitations in our simulations and the modeling of various systematic effects, there may be a mismatch between the simulated data and real observations. This simulation-observation mismatch, or distribution shift, can introduce significant biases in parameter inference \cite{villaescusa2021multifield}. In the second phase of the competition, participants have to tackle this problem by developing methods for OoD detection, i,e., identifying the test data that deviates from the training distribution. The main task is to construct a function $t(\boldsymbol{x})$ that assigns each input sample $\boldsymbol{x}$ a continuous OoD score, where larger values of $t(\boldsymbol{x})$ indicate a higher confidence that $\boldsymbol{x}$ is OoD. Ideally, $t(\boldsymbol{x})$ should be monotonically correlated with the confidence of distributional deviation.

In this phase, a subset of the test data is generated under different physical assumptions and is therefore not drawn from the training distribution. Participants are not provided with any OoD examples nor information about the underlying generative process.

\section{Metrics}
\label{sec:metrics}
Participants must submit their model predictions to the Codabench platform we set up. The submissions are then ranked by our predefined metrics on Codabench.

\subsection{Phase 1: Cosmological parameter estimation and uncertainty quantification}
In the first phase, participants' models should determine the point estimates $\hat{\Omega}_m$, $\hat{S}_8$, and their uncertainties $\hat{\sigma}_{\Omega_m}$, $\hat{\sigma}_{S_8}$. The model performance is ranked with the following score:
\begin{align}
    \text {score}_{\text{phase 1}} & =-\frac{1}{N_{\text {test }}} \sum_{i}^{N_{\text {test }}}\left\{\frac{\left(\hat{\Omega}_{m,i}-\Omega_{m,i}^{\text {truth}}\right)^2}{\hat{\sigma}_{\Omega_m,i}^2}+\frac{\left(\hat{S}_{8,i}-S_{8,i}^{\text {truth }}\right)^2}{\hat{\sigma}_{{S_8,i}}^2}\right. \nonumber \\
    & \left.\quad+\log \left(\hat{\sigma}_{\Omega_m,i}^2\right)+\log \left(\hat{\sigma}_{S_8,i}^2\right)+\lambda\left[\left(\hat{\Omega}_{m,i}-\Omega_{m,i}^{\text {truth }}\right)^2+\left(\hat{S}_{8,i}-S_{8,i}^{\text {truth }}\right)^2\right]\right\}.~\label{eq:inf_score}
\end{align}
The first term corresponds to the Kullback–Leibler (KL) divergence (up to some constants) between the true posterior distribution and the Gaussian distribution with the predicted mean and standard deviation. We expect the posterior distribution to be pretty Gaussian and the correlation between $\Omega_m$ and $S_8$ to be small, thus the Gaussian approximation with a diagonal covariance matrix should be good enough. The second term is an MSE loss with weight $\lambda=10^3$ to penalize bad point estimates.\footnote{The MSE penalty term in Eq.~(\ref{eq:inf_score}) can further be normalized with the prior ranges of $\Omega_{m}$ and $S_8$ to minimize the difference between the MSE scales of the two parameters of interest.}

\subsection{Phase 2: Out-of-distribution detection}
In the second phase, given a test sample $\boldsymbol{x}_i$, participants' models should give a continuous OoD score $t(\boldsymbol{x}_i)$ that increases monotonically with the confidence that their models predict a given sample as OoD.

\begin{figure}[t]
    \centering
    \includegraphics[width=0.8\linewidth]{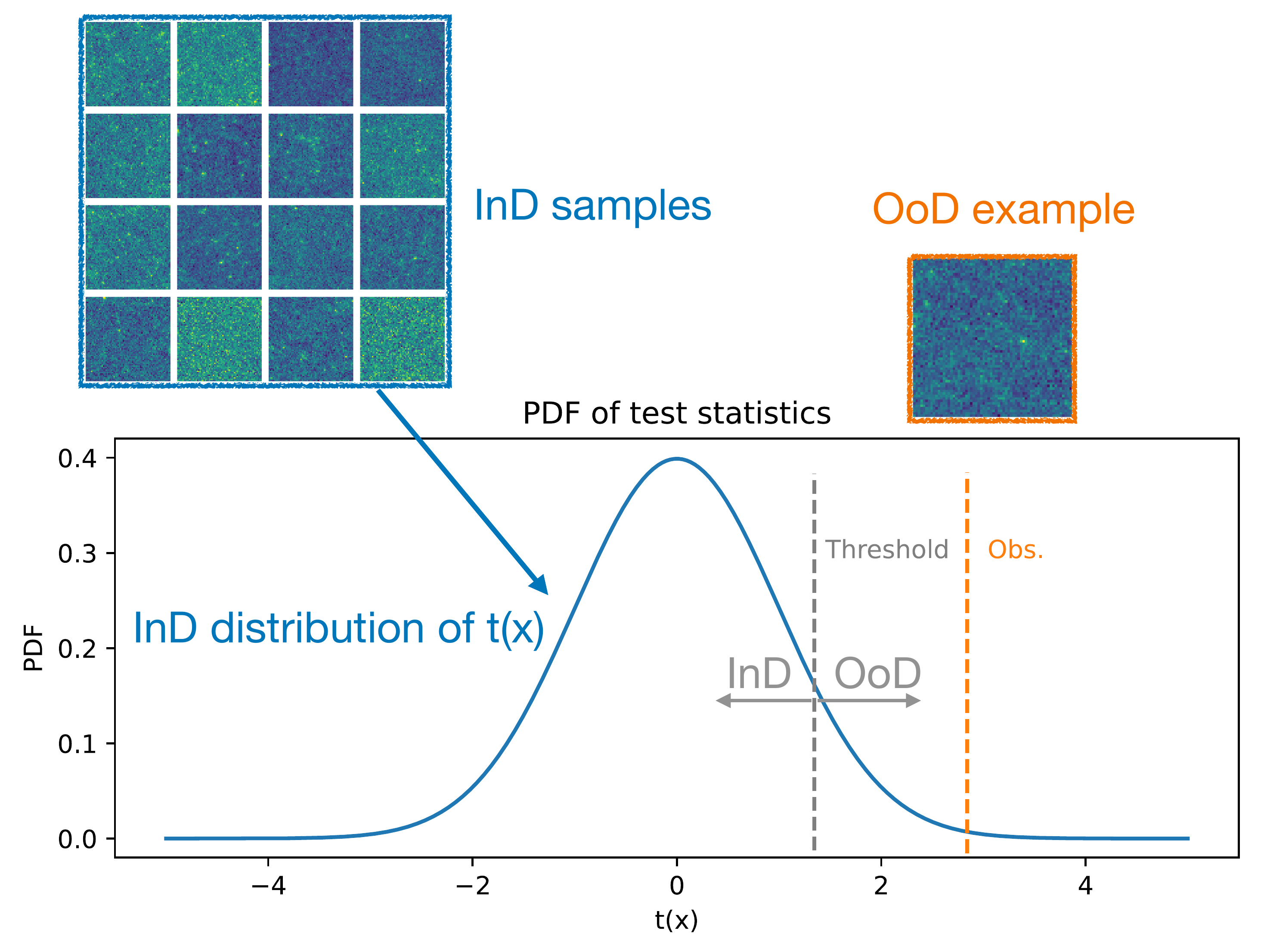}
    \caption{OoD detection with an OoD score $t(\boldsymbol{x})$ given test samples $\boldsymbol{x}$. \textit{OoD instances are detected if $t(\boldsymbol{x})$ is greater than a predefined threshold (a tunable parameter).} The figure is adapted from Ref.~\cite{Diao:2025szg}.}
    \label{fig:p2_ood_detection}
    \vspace{-0.2cm}
\end{figure}

Figure~\ref{fig:p2_ood_detection} shows that, by setting a threshold over the OoD score, samples with $t(\boldsymbol{x})$ above the threshold are classified as OoD, while those below are treated as InD. As the threshold varies, the predictions will lead to different true positive rates (TPRs, correctly identifying OoD samples) and false positive rates (FPRs, misclassifying InD samples as OoD). Tracing the pairs of FPR and TPR across all thresholds naturally defines the Receiver Operating Characteristic (ROC) curve. The OoD score $t(\boldsymbol{x})$ could be, for example, any test statistic increasing monotonically with the OoD likelihood, or the negative $p$-value defined from the test statistics of the training data and test data. 

The model's OoD detection performance will be evaluated by the mean values of the ROC curve over $N=100$ logarithmically spaced FPRs between $0.001$ and $0.05$; that is,
\begin{equation}
        {\rm score_{\text{phase 2}}} = \frac{1}{N}\sum_{i}^{N} \text{TPR}(\text{FPR}_i)  ~. \label{eq:ood_score}    
\end{equation}
This metric is approximately proportional to the area under the ROC curve over the given FPR range in logarithmic scale. 

The FPR range $[0.001,0.05]$ is chosen to match the regime of practical interest for scientific anomaly detection, where the FPR corresponds to the Type-I error rate (significance level $\alpha$). The interval spans thresholds from weak evidence ($\alpha \sim 0.05$) to stringent detection ($\alpha \sim 0.001$, approximately $3\sigma$). Focusing on this range therefore rewards models with high detection power under practically meaningful false-positive constraints, while logarithmic spacing emphasizes performance at the smallest FPRs.

\section{Baseline methods}

We provide several starting kit notebooks on the challenge website. The starting kits include simple scripts for loading and visualizing data, training and evaluating baseline methods, and preparing a submission file for the competition. The starting kits are designed to enable participants to easily get familiar with the dataset, the traditional methods for the challenge tasks, and the submission workflow. The baseline methods are used to set the baseline scores for the competition.

\subsection{Phase-1 baseline methods}
\label{sec:P1_baseline}
The baseline methods for the first phase of the competition include binning the power spectrum (a two-point correlation function of matter density fluctuations commonly used in cosmology) as a summary statistic or training a convolutional neural network to compress the data for downstream tasks or direct predictions.

\subsubsection{Power spectrum analysis with MCMC}
\label{sec:PS_MCMC}

In cosmology, the power spectrum describes how matter is distributed across different size scales in the universe and is a key tool for studying the growth of cosmic structure. Starting from the convergence map $\boldsymbol{x}(\boldsymbol{\theta})$, where $\boldsymbol{\theta}$ is the angular coordinate of the observed sky patch, we transform it into Fourier space to get 
\begin{equation}
    \tilde{\boldsymbol{x}}(\boldsymbol{k})
    = \int d^2\boldsymbol{\theta} \, 
      \boldsymbol{x}(\boldsymbol{\theta}) \,
      e^{-i\boldsymbol{k} \cdot \boldsymbol{\theta}}~,
\end{equation}
which represents fluctuations as waves of different wavelengths. The matter power spectrum $P_k$ is then defined by:
\begin{equation}
    \langle \tilde{\boldsymbol{x}}(\mathbf{k}) \tilde{\boldsymbol{x}}^*(\mathbf{k}') \rangle = (2\pi)^2 \delta_D(\mathbf{k}-\mathbf{k}') P_k~,
\end{equation}
where $k = |\mathbf{k}|$ is the wavenumber corresponding to a scale $\lambda \sim 1/k$, and $ \delta_D$ is the Dirac delta function. We can estimate the power spectrum by averaging over all Fourier modes within a wavenumber bin
\begin{equation}
    P_k = 
    \left\langle |\tilde{\boldsymbol{x}}(\boldsymbol{k})|^2 \right\rangle_{k \in \mathrm{bin}}~. \label{eq:Pk}
\end{equation}
Intuitively, $P_k$ tells us how ``clumpy'' the universe is on different scales. In cosmology, the shape and amplitude of $P_k$ encode the physics and composition of the universe, making it one of the most important statistical tools in the field. 

Given a convergence map generated from a simulation with cosmological parameters $\boldsymbol{\Theta} = (\Omega_m, S_8)$, the likelihood of an observation represented by a summary statistic $\boldsymbol{d}_{\rm obs}$ can be modeled as a Gaussian likelihood
\begin{equation}
    p(\boldsymbol{d}_{\rm obs}|\boldsymbol{\Theta}) \propto \frac{1}{\sqrt{|{\rm Cov}(\boldsymbol{\Theta})|}} \exp \left\{-\frac{1}{2}[\boldsymbol{d}_{\rm obs}-\mu(\boldsymbol{\Theta})]^T {\rm Cov}^{-1}(\boldsymbol{\Theta})[\boldsymbol{d}_{\rm obs}-\mu(\boldsymbol{\Theta})]\right\}~.     \label{eq:gauss_lh}
\end{equation}
When there are $N_{\rm sys}$ realizations (parameterized by nuisance parameters $\boldsymbol{\alpha}$) for each cosmological model, the mean summary statistic evaluated at $\boldsymbol{\Theta}$ is
\begin{equation}
    \mu(\boldsymbol{\Theta}) =\frac{1}{N_{\rm sys}} \sum_{j=1}^{N_{\rm sys}} \boldsymbol{d}(\boldsymbol{\Theta},\boldsymbol{\alpha}_j)~, \label{eq:mu}
\end{equation}
where $\boldsymbol{d}(\boldsymbol{\Theta},\boldsymbol{\alpha}_j)$ is the summary statistic of the $j$-th realization of cosmology $\boldsymbol{\Theta}$, and the covariance matrix of the summary statistic can be estimated with
\begin{equation}
    \operatorname{Cov}(\boldsymbol{\Theta})=\frac{1}{N_{\rm sys}-n_d-2} \sum_{j=1}^{N_{\rm sys}}~[\boldsymbol{d}(\boldsymbol{\Theta},\boldsymbol{\alpha}_j)-\mu(\boldsymbol{\Theta})]^T ~[\boldsymbol{d}(\boldsymbol{\Theta},\boldsymbol{\alpha}_j)-\mu(\boldsymbol{\Theta})]~,    \label{eq:cov}
\end{equation}
where $n_d$ is the number of summary statistics.

In the standard power spectrum analysis, the logarithm of the power spectrum in Eq.~(\ref{eq:Pk}) is treated as a summary statistic for Eq.~(\ref{eq:gauss_lh}) to estimate the cosmological parameters
\begin{equation}
    \boldsymbol{d}(\boldsymbol{\Theta}, \boldsymbol{\alpha}) = \text{log} ~P_k(\boldsymbol{\Theta}, \boldsymbol{\alpha})~,  \label{eq:summary_statistic_PS}
\end{equation}
where $n_d$ is the number of bins in wavenumber $k$. 

The power spectrum emulator $P_k(\boldsymbol{\Theta}, \boldsymbol{\alpha})$ can be obtained using the training dataset we provide. At the inference time, we sample the posterior $p(\boldsymbol{\Theta}|\boldsymbol{d}_{\rm obs})$ with MCMC using the likelihood function (\ref{eq:gauss_lh}) and a flat prior of $\boldsymbol{\Theta}$, and compute the the point estimates $\hat{\boldsymbol{\Theta}} = (\hat{\Omega}_m$, $\hat{S}_8)$ and their uncertainties $(\hat{\sigma}_{\Omega_m}$, $\hat{\sigma}_{S_8})$ from the posterior samples. 

In our starting kit, we also show that the OoD detection probability can be conservatively estimated by comparing the dataset likelihood with the distribution of the training data likelihood.

\subsubsection{Convolutional neural network with MCMC}
\label{sec:CNN_MCMC}

Neural networks are very powerful in extracting useful features from high-dimensional data. In this baseline method, we model the mapping from convergence maps to the cosmological parameters $(\Omega_m, S_8)$ using a convolutional neural network (CNN) implemented in \textsc{PyTorch}~\cite{pytorch}.

The input of the CNN is a single–channel 2D convergence map of $1424 \times 176$ pixels. The architecture consists of four convolutional blocks, each composed of a 2D convolution, batch normalization, ReLU activation, and $2\times 2$ max pooling. The numbers of feature channels are $16$, $32$, $64$, and $128$ for the four convolutional blocks, respectively. The first convolution uses a $5\times 5$ kernel with stride $2$ and padding $2$, while the remaining convolutions use $3\times 3$ kernels with stride $1$ and padding $1$.

The output of the convolutional stack is then flattened and passed through a fully connected regression head: two linear layers with $512$ and $128$ units, each followed by ReLU activation and dropout with probabilities $0.2$ and $0.1$, respectively, and a final linear layer outputting the two target parameters.

The CNN is trained using mean–squared error (MSE) loss between the predicted and true $(\Omega_m, S_8)$ values. Optimization is performed using \texttt{Adam} with a learning rate of $2\times 10^{-4}$, weight decay of $10^{-4}$, batch size of $64$, and for a total of $15$ epochs. We apply a \texttt{ReduceLROnPlateau} scheduler that decreases the learning rate by a factor of $0.5$ if the validation loss does not improve for five consecutive epochs. The training is performed using GPUs when available.

The outputs of the trained CNN are the point estimates $\hat{\boldsymbol{\Theta}} = (\hat{\Omega}_m, \hat{S}_8)$, which can be treated as the summary statistic for Eq.~(\ref{eq:gauss_lh}) 
\begin{equation}
    \boldsymbol{d}(\boldsymbol{\Theta}, \boldsymbol{\alpha}) = f_{\rm NN}^{\phi}\left(\boldsymbol{x}(\boldsymbol{\Theta}, \boldsymbol{\alpha})\right) =\hat{\boldsymbol{\Theta}} = (\hat{\Omega}_m, \hat{S}_8)~,  \label{eq:summary_statistic_CNN}
\end{equation}
where $n_d = 2$. The MCMC is then applied to generate samples of the posterior distribution that can be used to compute the uncertainties.

\begin{figure}[t]
    \centering
    \includegraphics[width=0.8\linewidth]{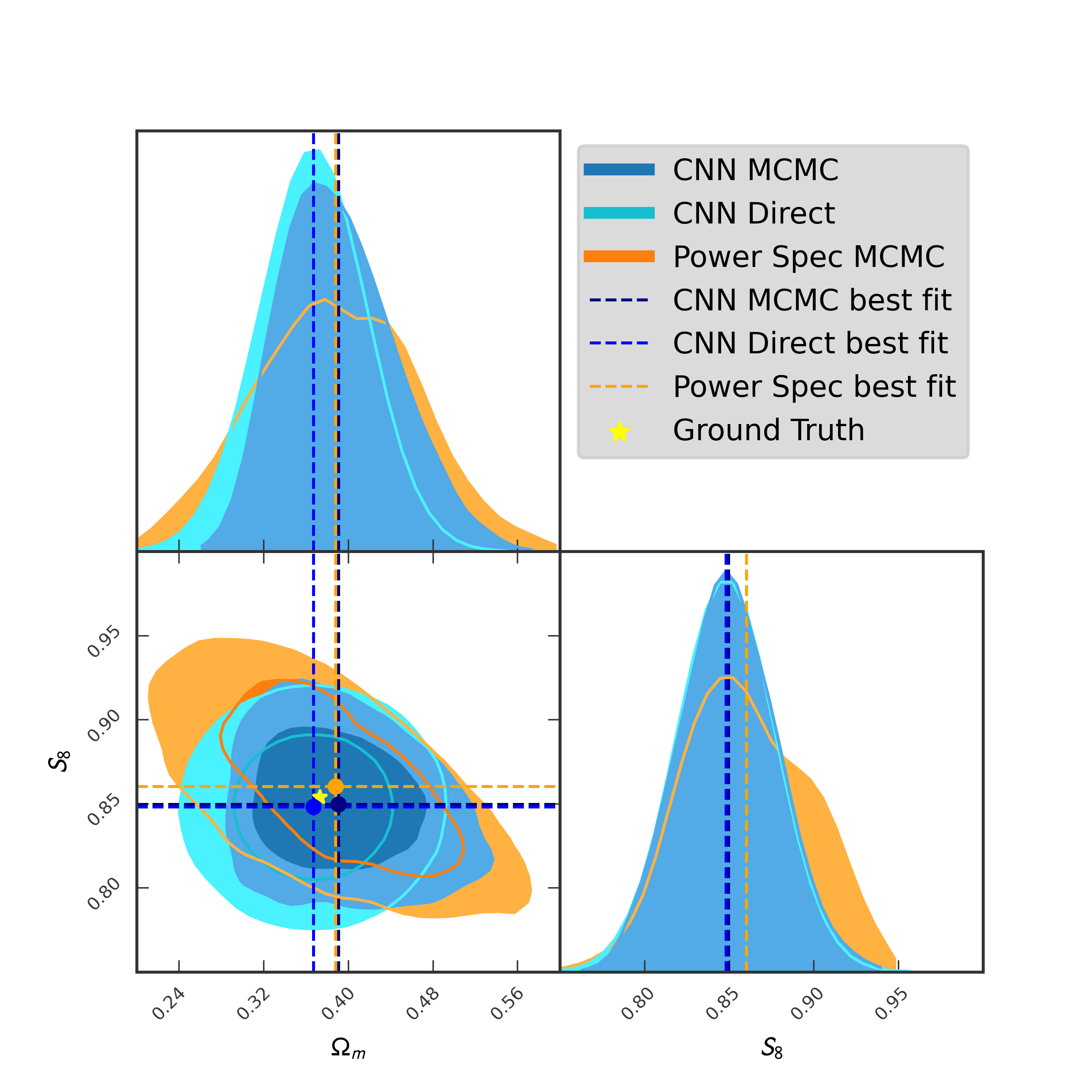}
    \caption{A comparison between the posterior distributions inferred by our baseline methods. \textit{The neural network-based approaches can  estimate the parameters more accurately.}}
    \label{fig:p1_baseline_comp}
    \vspace{-0.2cm}
\end{figure}

\subsubsection{Convolutional neural network direct prediction}
\label{sec:CNN_dir}

In this baseline method, we do not fit any predefined summary statistic to the data and perform MCMC. Instead, we estimate the uncertainties directly using the CNN. This is achieved by optimizing a KL divergence objective function using neural network predictions during training
\begin{equation}
    \mathcal{L}_{\rm KL} = \frac{1}{N} \sum_i^{N}\left\{\frac{\left(\hat{\Omega}_{m, i}-\Omega_{m, i}^{\text {truth }}\right)^2}{\hat{\sigma}_{\Omega_m, i}^2}+\frac{\left(\hat{S}_{8, i}-S_{8, i}^{\text {truth }}\right)^2}{\hat{\sigma}_{S_8, i}^2}+\log \left(\hat{\sigma}_{\Omega_m, i}^2\right)+\log \left(\hat{\sigma}_{S_8, i}^2\right)\right\}~.    
\end{equation}

We use the same CNN architecture and hyperparameters described in Sec.~\ref{sec:CNN_MCMC}, except that we change the output dimension of the final layer to $4$ so that for each 2D convergence map the CNN will predict its cosmological parameters $(\hat{\Omega}_m, \hat{S}_8)$ and the standard deviations of the joint Gaussian posterior distribution $(\hat{\sigma}_{\Omega_m}, \hat{\sigma}_{S_8})$. The advantage of this approach is that it directly forces the CNN to learn to predict the point estimates and uncertainties optimizing the scoring function defined in Eq.~(\ref{eq:inf_score}) while avoiding choosing any summary statistic and saving lots of time from MCMC sampling.

\subsubsection{Baseline scores}
Figure~\ref{fig:p1_baseline_comp} shows a comparison between the sampled posterior distributions of our three baseline methods. Using the scoring metric defined in Eq.~(\ref{eq:inf_score}), the power spectrum analysis, CNN with MCMC, and CNN direct prediction obtain scores of $4.5766$, $8.6809$, and $8.5209$, respectively, on the Phase-1 public test dataset. Compared to the traditional power spectrum analysis, the neural networks can capture more information from the weak lensing data, leading to better predictions.

\subsection{Phase-2 baseline methods}
\label{sec:P2_baseline}
The baseline methods for the second phase of the competition include estimating the $p$-values by comparing $\chi^2$ values between test and training data, following the Phase-1 baseline methods, or by comparing reconstruction errors between OoD and InD samples using a trained autoencoder.

\subsubsection{Chi-squared distributions from the Phase-1 baseline methods}
With the baseline MCMC methods of Phase 1, we assess the goodness-of-fit of the test sample $i$ using the $\chi^2$ statistic
\begin{equation}
    \chi_i^2 = [\boldsymbol{d}_i-\mu(\hat{\boldsymbol{\Theta}}_i)]^T {\rm Cov}^{-1}(\hat{\boldsymbol{\Theta}}_i)[\boldsymbol{d}_i-\mu(\hat{\boldsymbol{\Theta}}_i)]~,     \label{eq:gauss_lh}
\end{equation}
where $\mu(\hat{\boldsymbol{\Theta}}_i)$ and ${\rm Cov}(\hat{\boldsymbol{\Theta}_i})$ are defined by Eqs.~(\ref{eq:mu}) and (\ref{eq:cov}), respectively, and are estimated at the best-fit parameters $\hat{\boldsymbol{\Theta}}_i = (\hat{\Omega}_{m,i}, \hat{S}_{8,i})$. The summary statistic $\boldsymbol{d}_i$ is given by Eqs~(\ref{eq:summary_statistic_PS}) or (\ref{eq:summary_statistic_CNN}). 

We then estimate the $p$-value for the test sample $i$ using the $\chi^2$ distribution of the training set that is only composed of InD samples by computing the fraction of InD training samples whose $\chi^2$ values are smaller than the test realization
\begin{equation}
    p\text{-value}_i = \text{probability}(\chi_{\rm train~(InD)}^2 > \chi_i^2)~.
\end{equation}
We take the negative $p$-value as the OoD score. Our baseline predictions are then scored by Eq.~(\ref{eq:ood_score}).

\subsubsection{Reconstruction errors with autoencoder}
An alternative approach to estimate the InD probability is to quantify how well the test convergence maps can be reconstructed by a neural network trained exclusively on the InD samples. We train a convolutional autoencoder (AE) on the training set, which contains only InD realizations drawn from the fiducial cosmological prior. The AE learns a low-dimensional latent representation $\boldsymbol{z}$ of each convergence map $\boldsymbol{x}$ through an encoder $\mathcal{E}_{\rm NN}^{\phi}(\boldsymbol{x})$, and reconstructs the input via a decoder $\mathcal{D}_{\rm NN}^{\phi}(\boldsymbol{z})$. The reconstructed map is given by
\begin{equation}
    \hat{\boldsymbol{x}} = \mathcal{D}_{\rm NN}^{\phi}(\boldsymbol{z} = \mathcal{E}_{\rm NN}^{\phi}(\boldsymbol{x}))~. 
\end{equation}

For the network architecture, we employ a convolutional autoencoder with a low-dimensional latent bottleneck to quantify the degree to which each convergence map can be faithfully reconstructed by a model trained exclusively on InD data. The encoder consists of two convolutional layers with stride-2 downsampling and ReLU activations, mapping an input map of size $1424\times176$ to a feature map of size $32\times356\times44$. A dropout layer with a rate of $0.1$ is applied in the encoder to regularize the learned representation. The resulting feature map is flattened and projected to an $8$-dimensional latent representation $\boldsymbol{z}$ through a fully connected layer. In the decoder, the latent vector is first projected back to the flattened convolutional feature space and reshaped into a $32\times356\times44$ tensor, which is then upsampled by two transposed-convolution layers to reconstruct the map. The optimization is performed using \texttt{Adam} with a learning rate of $2\times 10^{-4}$, weight decay of $10^{-4}$, batch size of $64$, and for a total of $15$ epochs.

For simplicity, we train the model using a purely reconstruction-based objective. Specifically, the loss function minimizes the mean-squared reconstruction error
\begin{equation}
    \mathcal{L}_{\rm recon} = \|\boldsymbol{x} - \hat{\boldsymbol{x}}\|^2 ~, \label{eq:loss_recon}
\end{equation}
while the Kullback–Leibler regularization term normally used in VAEs is set to zero. As a result, the network behaves as a deterministic autoencoder during training, learning a compressed representation that preserves only the information required to reconstruct InD maps accurately. Mathematically, minimizing Eq.~(\ref{eq:loss_recon}) is equivalent to maximizing the log-likelihood under the assumption that the data errors or residuals follow an independent and identically distributed Gaussian distribution with a fixed variance.

For each map $\boldsymbol{x}_i$, the autoencoder produces a reconstruction $\hat{\boldsymbol{x}}_i$, and the reconstruction error is calculated by Eq.~(\ref{eq:loss_recon}). The distribution of reconstruction errors obtained from the training set provides an empirical reference for InD data. The test samples that yield significantly large or small reconstruction errors relative to the training distribution are more likely to be OoD realizations. We directly calculate the $p$-values by comparing the test and the training errors.

\begin{figure}[t]
    \centering
    \begin{subfigure}[t]{0.5\textwidth}
        \centering
        \includegraphics[width=\linewidth]{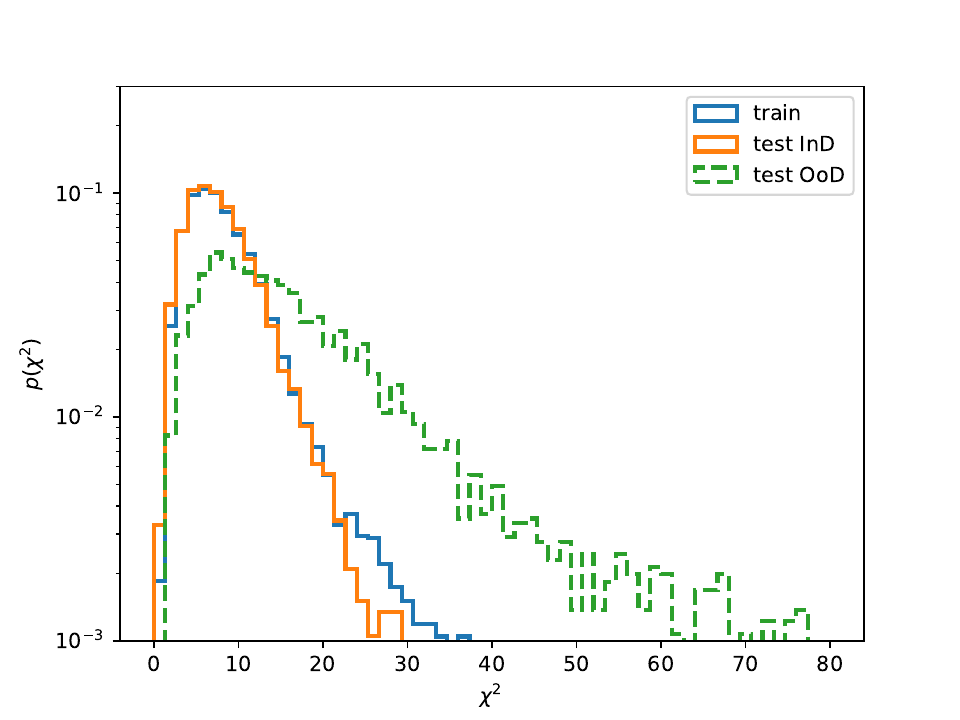}
        \caption{\textbf{Power Spectrum Analysis:} $\chi^2$ distributions}
    \end{subfigure}\hspace{-4mm}
    \begin{subfigure}[t]{0.5\textwidth}
        \centering
        \includegraphics[width=\linewidth]{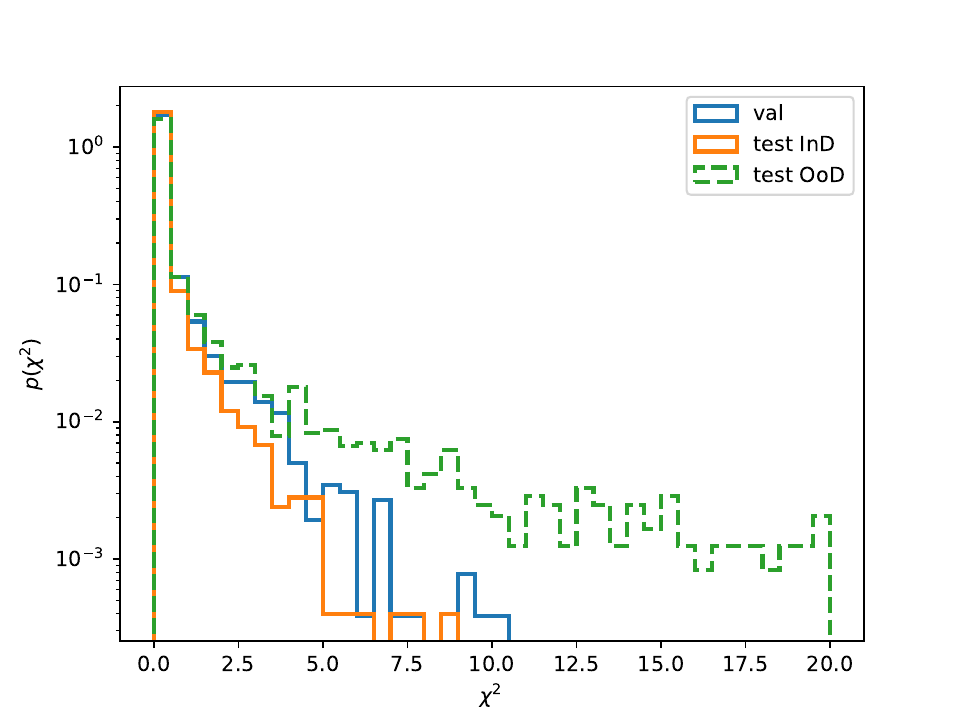}
        \caption{\textbf{CNN with MCMC:} $\chi^2$ distributions}
    \end{subfigure}\hspace{-4mm}
    \begin{subfigure}[t]{0.5\textwidth}
        \centering
        \includegraphics[width=\linewidth]{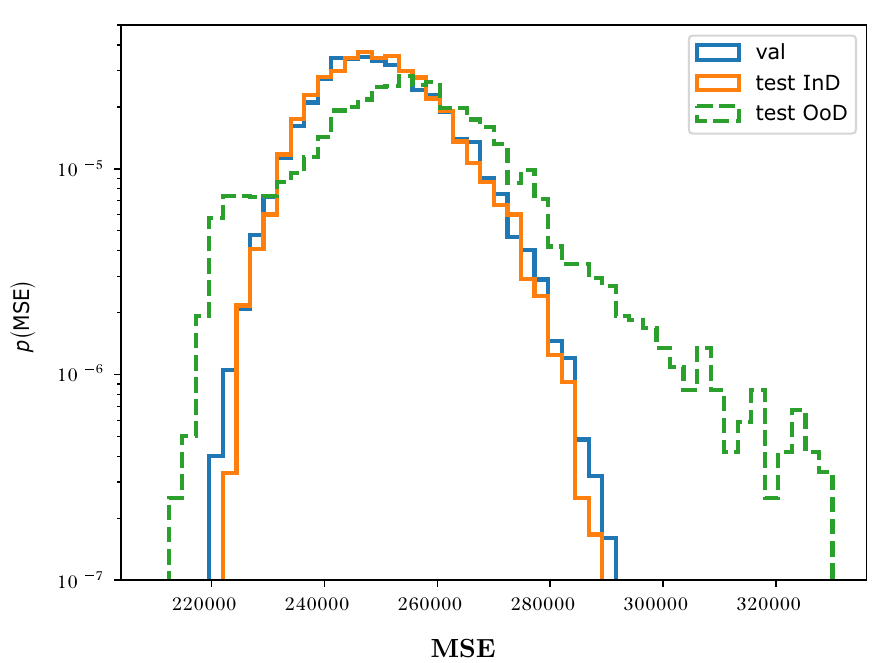}
        \caption{\textbf{Autoencoder:} Reconstruction error distributions}
    \end{subfigure}
    \caption{A comparison between all baseline methods for the Phase-2 task. \textit{The InD and OoD samples in the test data can be partially distinguished}.}
    \label{fig:p2_baseline_comp}
\end{figure}

\subsubsection{Baseline scores}
Figure~\ref{fig:p2_baseline_comp} shows the $\chi^2$ distributions from the power spectrum analysis and the CNN MCMC method, as well as the distribution of reconstruction errors from the autoencoder. Using the scoring metric defined in Eq.~(\ref{eq:ood_score}), the power spectrum analysis, CNN, and autoencoder obtain leaderboard scores of $0.2143$, $0.1053$, and $0.1307$, respectively, on the Phase-2 public test dataset. Random OoD scores will give a leaderboard score of $0.0128$. We find that simple neural network-based baseline methods do not perform as well as the power spectrum analysis across some OoD categories in the test data. Participants have to enhance the baseline solutions so that the models learn a better representation of the data to detect those OoD categories.

\section{Summary}

Through this challenge, we have established the first benchmark dataset for ML applications in weak gravitational lensing cosmology. A key feature of our simulated dataset is the inclusion of realistic systematic effects, such as the baryonic feedback and the photometric redshift uncertainty, enabling the community to develop and evaluate methods that could be generalized to the real data of the upcoming cosmological surveys.

The goals of the challenge are two-fold: (i) to accurately infer key cosmological parameters from high-dimensional weak lensing maps while providing reliable uncertainty estimates; and (ii) to detect and quantify distribution shifts arising from imperfect simulations or mismatched physical models through the OoD detection. By structuring the competition into two complementary phases, we provide a unified framework for benchmarking both parameter inference and robustness to modeling errors within the same dataset.

We presented several baseline methods spanning traditional power spectrum analyses and modern neural network-based approaches, illustrating clear performance gains from machine-learning models in extracting non-Gaussian information, while also highlighting the challenges posed by OoD scenarios. The baseline results demonstrate that, although neural networks can significantly improve cosmological parameter estimation in-distribution, robust uncertainty quantification and OoD detection remain open problems that require further methodological advances.

The challenge is designed to foster close collaboration between the cosmology and machine-learning communities, encouraging the development of uncertainty-aware inference techniques for weak lensing cosmology. We anticipate that the dataset, metrics, and baseline implementations introduced here will serve as a lasting reference for future research and will help inform the design of next-generation weak lensing analysis pipelines for upcoming surveys such as Euclid, the Vera Rubin Observatory, and the Nancy Grace Roman Space Telescope.

\section*{Acknowledgments}
We are grateful to the US Department of Energy, Office of High Energy Physics, subprogram on Computational High Energy Physics, for sponsoring this research. BD acknowledges support from the Ambrose Monell Foundation, the Corning Glass Works Foundation Fellowship Fund, and the Institute for Advanced Study. This research used resources of the National Energy Research Scientific Computing Center (NERSC), a Department of Energy Office of Science User Facility, using NERSC award HEP-ERCAP0032917.

\bibliographystyle{tepml}
\bibliography{ref.bib,HEPML}



\end{document}